\documentclass[]{aastex}
\usepackage{emulateapj5}
\input{epsf.sty}

\def\ifm#1{\relax\ifmmode#1\else$\mathsurround=0pt #1$\fi}
\def\Mmin{{\rm M}_{\rm min}}
\def\Mbar{\langle{\rm M}\rangle}
\def\Mn{{\rm M}_{1}} 
\def\Ng{{N}_{g}} 
\def\hoh{h$_{100}^{-1}$}
\def\hmpc{\,{h$_{100}^{-1}$\,Mpc}}
\def\msun{{\rm M}_{\odot}}
\def\hmsun{{\rm h}_{100}^{-1}\msun}
\def\lcdm{$\Lambda$CDM}
\def\ltsima{$\; \buildrel < \over \sim \;$}
\def\lsim{\lower.5ex\hbox{\ltsima}}
\def\gtsima{$\; \buildrel > \over \sim \;$}
\def\gsim{\lower.5ex\hbox{\gtsima}}
\def\xidm{\xi_{\rm DM}(r)}

\newenvironment{inlinefigure}{
\def\@captype{figure}
\noindent\begin{minipage}{0.999\linewidth}\begin{center}}
{\end{center}\end{minipage}\smallskip}

\lefthead{MOUSTAKAS \& SOMERVILLE}
\righthead{SPATIAL CLUSTERING EVOLUTION OF EXTREMELY RED OBJECTS}

\begin{document}

\submitted{Accepted for publication in the Astrophysical Journal}

\title{The Masses, Ancestors and Descendents of Extremely Red Objects: \\
Constraints from Spatial Clustering}

\medskip
\author{Leonidas~A.~Moustakas}
\affil{Nuclear and Astrophysics Laboratory, University of Oxford, Keble
Road, OX1\,3RH, UK\\ {\tt leonidas@astro.ox.ac.uk}}
\medskip
\and
\author{Rachel~S.~Somerville}
\affil{Astronomy Department, University of Michigan, Ann Arbor, MI 
48109\\{\tt rachel@astro.lsa.umich.edu}} 
\medskip

\begin{abstract}
Wide field near-infrared (IR) surveys have revealed a population of
galaxies with very red optical$-$IR colors, which have been termed
``Extremely Red Objects'' (EROs). Modeling suggests that such red
colors ($R-K > 5$) could be produced by galaxies at $z \gsim 1$ with
either very old stellar populations or very high dust extinction.
Recently it has been discovered that EROs are strongly clustered. Are
these objects the high-redshift progenitors of present day giant
ellipticals (gEs)?  Are they already massive at this epoch?  Are they
the descendents of the $z\sim3$ Lyman Break Galaxies (LBG), which have
also been identified as possible high redshift progenitors of giant
ellipticals?  We address these questions within the framework of the
Cold Dark Matter paradigm using an analytic model that connects the
number density and clustering or bias of an observed population with
the halo occupation function (the number of observed galaxies per halo
of a given mass). We find that EROs reside in massive dark matter
halos, with average mass
$\Mbar>10^{13}$\,h$_{100}^{-1}$\,M$_{\odot}$. The occupation function
that we derive for EROs is very similar to the one we derive for
$z=0$, $L>L_*$ early type galaxies, whereas the occupation function
for LBGs is skewed towards much smaller host halo masses ($\Mbar
\approx 10^{11-12}$\,h$_{100}^{-1}$\,M$_{\odot}$).  We then use the
derived occupation function parameters to explore the possible
evolutionary connections between these three populations.
\end{abstract}

\begin{keywords}{galaxies: high-redshift --- galaxies: halos ---
galaxies: evolution --- galaxies: fundamental parameters --- theory:
dark matter}

\end{keywords}

\section{Introduction\label{sec:intro}}

The first generation of deep near-IR surveys turned up a population of
objects not represented in optical surveys (\citealt*{elston:88};
\citealt*{elston:89}; \citealt*{mcc:92}; \citealt*{hr:94};
\citealt{cowie:94}; \citealt{mous:97}; \citealt{barger:99};
\citealt{thompson:99}). These objects are remarkable for their very
red optical$-$infrared colors (e.g. $R-K>6$), and hence gained the
moniker ``Extremely Red Objects'' (EROs). Using either empirical
templates or stellar population modeling, it is found that these
extreme colors could be produced in galaxies at $z \gsim 1$ with
either a very old and quiescent stellar population or very large dust
extinction \citep[see e.g.][Fig.~1]{firth:02}.  At these redshifts,
the observed magnitudes $K \lsim 19$ imply large luminosities, $L >
L_*$.  Taken together, these clues hint that EROs could be the elusive
high redshift progenitors of present-day giant ellipticals, already
massive and with evolved stellar populations at an epoch when the
total age of the Universe is only $4-5$\,Gyr. If correct, this clearly
has dramatic implications for theories of galaxy formation. In
particular, it casts in sharp contrast the hierarchical, Cold Dark
Matter (CDM) based scenario, in which ellipticals form in gas-rich
mergers, and the (phenomenological) ``monolithic collapse'' scenario
\citep*{els:62}, in which ellipticals form at high redshift and evolve
passively.

For several years, further progress in the interpretation of these
objects was hindered by the difficulty of obtaining reliable
measurements of their number densities and redshifts. The surface
densities of EROs measured in the early fields varied widely
\citep[e.g.][]{cowie:94, mous:97, thompson:99}, perhaps not
surprisingly in view of the small volumes that had been probed. In
addition, the faintness and apparent lack of dramatic spectral
features has made a systematic spectroscopic study of a large sample
of EROs impractical. Heroic spectroscopic efforts directed at a few
objects secured several high signal-to-noise-ratio spectra,
demonstrating that some of the EROs indeed contain predominantly old
stellar populations with very little recent star formation and are at
redshifts of $z\sim1$ \citep{spinrad:97, dunlop:96}. Some, however,
appear to be heavily dust enshrouded starburst galaxies, similar to
local ULIRGs, also at redshifts $z\sim1$ \citep{gd:96, smail:99}. 

Recently, the completion of portions of several wide field surveys
with multi-band, optical to near-IR photometry
\citep[e.g.][]{daddi:00a, firth:02, chen:02, olding:01, martini:01}
has led to several interesting developments. The area surveyed is
finally large enough that the global average surface density of EROs
is more robustly measured (though still with some uncertainty) to
limiting magnitudes $K\approx 19-20$. These surveys have also shown
that the EROs are strongly clustered on the sky, with an angular
clustering amplitude $\approx10$ times larger than the overall
population at the same limiting magnitude. To obtain estimates of the
real-space clustering, one needs redshift information.  Obtaining
large numbers of spectroscopic redshifts has proven to be extremely
difficult, but photometric redshifts indicate that EROs selected at,
for example, $I-H>3$, lie in a relatively narrow redshift range with a
mean of $z \simeq 1.2$ \citep{firth:02}. This implies a real-space
correlation length of $r_0 \sim 9-10$\hmpc\ \citep[precise values
depending on the color and magnitude cut]{mcc:01, firth:02, daddi:01},
comparable to or larger than $z=0$ luminous early type galaxies.

Gradual progress is also being made in determining the break-down of
the ERO population in terms of the two primary classes of ``old'' and
``dusty'' objects, using judiciously chosen broadband photometry
\citep*[e.g.][]{pozzetti:00}, spectroscopy \citep{cimatti:02}, and
morphology \citep{moriondo:00}. From all of these measures, the
fraction of ``old'' objects is consistent with 50$\pm$20\%, although
the contribution from dusty objects may be larger at faint
magnitudes. Based on a small sample of 35 EROs with spectroscopic
redshifts, \citet{daddi:02} found that the ``dusty'' objects are
significantly less clustered than the ``old'' objects. The correlation
length for the old EROs is constrained to be in the range $r_0 \sim
5.5-16$\hmpc, consistent with previous estimates based on the full
photometric samples.

Knowledge about the clustering properties of these objects provides
important clues about their nature and their connection with other
populations. In the Cold Dark Matter (CDM) paradigm, massive halos
correspond to rare peaks in the density field and are therefore
strongly clustered. Thus we expect a correlation between halo mass and
clustering amplitude, and the description of this correlation by
analytic theories \citep{mw:96, st:99} agrees quite well with the
results of N-body simulations. If the values of cosmological
parameters are determined by other means, then in principle we can use
the observed clustering properties of any population to constrain the
possible masses of the host dark matter halos. In practice, an added
complication is the unknown ``occupation function'' of the observed
population; i.e. the number of observed galaxies per dark matter halo
as a function of halo mass.  For any number- or pair-weighted
statistic like the correlation function, this effectively gives
different weights to halos of different masses. The occupation
function in essence provides the link between an observed population
and the theoretically tractable population of dark matter halos.

The occupation function is determined, in reality, by the host of
interwoven astrophysical processes controlling gas cooling, star
formation, feedback, and so on. It may be computed from first
principles using detailed models of galaxy formation,
i.e. semi-analytic models \citep{somerville:01, benson:01b} or
hydrodynamic simulations \citep{whitem:01}. Numerous studies have run
these ``forward'' models to fill dark matter halos with galaxies, and
then evaluated whether the resulting number densities and clustering
properties match observations for various populations
(e.g. \citealt{wechsler:01}; \citealt*{spf:01}; \citealt*{khw:99}).
Alternatively, an empirical approach may be taken, which bypasses the
necessity of making any assumptions about the \emph{physics} of galaxy
formation.  If a specific functional form for the occupation function
is adopted, its parameters can be constrained by appealing directly to
observations.  For example, \citet{peacock:00} and \citet{marinoni:02}
do this using the observed luminosity function of groups, and
\citet*{berlind:01} use the clustering properties of nearby galaxies. 

In this paper, we make use of a similar empirical approach, using a
simple analytic model for galaxy clustering, based on the formalism
introduced in \citet{wechsler:01} and \citet*[][BWS02]{bullock:02}. In
this model, we parametrize the occupation function using a simple
functional form, a power law characterized by three parameters: a
minimum mass, a slope, and a normalization. This form may be motivated
by more detailed models of galaxy formation such as semi-analytic
models or hydrodynamic simulations \citep[see
e.g.][]{wechsler:01,benson:01b,whitem:01}.  Similar formalisms have
been presented by e.g., \citet{jing:98}, \citet{seljak:01},
\citet{scocc:01}, and \citet*{berlind:01}.  Simultaneous information
on the number density of a population as well as its clustering on
small and large scales can effectively constrain all three parameters
of the occupation function expressed in this way (BWS02). However,
measuring the small scale clustering requires accurate redshifts which
are not available for the bulk of the EROs. Therefore, here we treat
the slope of the occupation function as a free parameter and
investigate the allowed range of values for the minimum mass and
normalization, based on the constraints from the number density and
large scale clustering of EROs. We use similar constraints for local
giant ellipticals and the $z\sim3$ Lyman Break Galaxies (LBGs) to
derive their occupation functions, and use this information to
speculate on the relationship of these populations to the intermediate
redshift EROs.

This empirical approach provides a useful complement to the detailed,
physical models in several respects, perhaps particularly in the case
of high redshift objects such as the EROs. Semi-analytic models based
in the currently favored \lcdm\ cosmology, which do well at
reproducing the observed properties of $z=0$ and $z=3$ galaxies
\citep{sp:99,spf:01,cole:00}, fail to reproduce the observed number
densities of EROs, by a significant factor (\citealt{firth:02};
Somerville \& Moustakas, in prep).  The continuing uncertainty in the
basic nature of the bulk of EROs (i.e. whether ``old'' or ``dusty'')
makes it difficult to isolate the physics responsible for the
discrepancy. The results presented here provide direct constraints on
such models in terms of the basic theoretical quantities. Furthermore,
in the hierarchical paradigm, the age of the constituents of a system
(e.g. stars) does not necessarily bear any relationship to the epoch
of {\em assembly} of the object. Therefore, while numerous
observations point to the conclusion that the stars in elliptical
galaxies are rather uniformly old, this does not place any direct
constraint on the {\em mass assembly history} of these galaxies. The
clustering analysis that we present here circumvents this ambiguity,
again by directly constraining the masses of the dark matter halos
hosting the plausible progenitors of giant ellipticals at high
redshift.

The structure of the paper is as follows. In \S\ref{sec:data}, we
summarize the relevant observational data for the three populations
that we will study: local giant ellipticals (gEs), intermediate
redshift EROs, and high redshift LBGs. In \S\ref{sec:models}, we
provide a brief summary of the analytic model.  In \S\ref{sec:invert},
we present constraints on the occupation function parameters for the
three populations, and in \S\ref{sec:zev}, we present predictions for
the redshift evolution of the number density and clustering properties
of each population.  In \S\ref{sec:conclusions} we summarize our results
and main conclusions.

All computations are carried out for a $\Lambda$CDM cosmology, with
$\Omega_{\Lambda}=0.7$, $\Omega_0=0.3$, and a power-spectrum
normalization of $\sigma_8=0.9$. In calculating the power spectrum, we
assume H$_0=70$\,km\,s$^{-1}$\,Mpc$^{-1}$, but all results are scaled
to h$_{100}$\,$\equiv$\,H$_0/(100$\,km\,s$^{-1}$\,Mpc$^{-1}$).

\section{Observational Constraints}
\label{sec:data}

For any given population of interest, we require the comoving number
density and a measure of the clustering strength.  In general,
observed correlation functions are characterized by a power law
functional form,
\begin{equation}
\xi(r) = \left(\frac{r}{r_0}\right)^\gamma.
\end{equation}
We can then convert this to a {\em bias} value, where the bias
represents the clustering strength of the galaxies ($\xi_g$) with
respect to that of the dark matter ($\xi_{\rm DM}$) at a given epoch
and scale. Several different definitions of bias appear in the
literature; we use the square root of the ratio of the correlation
functions at 8\hmpc:
\begin{equation}
b \equiv (\xi_g/\xi_{\rm DM})^{1/2}.
\end{equation}

The correlation amplitude of the dark matter is straightforward to
obtain for an assumed cosmological model, using either numerical
simulations or analytic approximations.  We have computed convenient
fits for the dark matter correlation scale length $r_0$ and slope
$\gamma$ from the GIF/VIRGO N-body \lcdm\ simulation
\citep{jenkins:98}, which are described and shown in the Appendix.
The dark matter correlation scale as a function of redshift is shown
in Fig.~\ref{fig:zevol}.

We consider three observed populations:
\begin{itemize}
\item {\em Local giant elliptical galaxies} (gEs, $z\sim0$)
\item {\em Extremely Red Objects} (EROs, $z\sim1.2$)
\item {\em Lyman Break Galaxies} (LBGs, $z\sim3$)
\end{itemize}
The definition of each of these populations is somewhat arbitrary,
particularly in the case of the high redshift objects, about which not
very much is known. As we will rely on results from the literature, we
are restricted to definitions used there. While the values of the
number density and clustering statistics are inevitably sensitive to
the details of how the sample is selected (especially to the magnitude
limit), here we have chosen as a matter of philosophy not to address
the uncertainties due to the ambiguity of sample selection. Rather, we
quote uncertainties that reflect the {\em observational} uncertainty
associated with the determination of the relevant quantity {\em for a
given sample}. These values should be considered representative only.
The results should be updated as better observational data become
available.

Nearby ``early type'' galaxies may be defined in a number of different
ways, e.g. by morphological type, color, or spectral type. We wish to
focus on {\em luminous} early type galaxies, as the EROs with which we
want to compare them correspond to an intrinsically luminous
population (if they are at the assumed redshifts $z\sim1$). Locally,
early type and red galaxies are well-known to be a factor of about
1.2--1.5 times more strongly clustered than the overall population at
a given magnitude limit. Moreover, brighter galaxies also seem to be
more strongly clustered than less luminous ones. However, as the
bright end of the luminosity function becomes increasingly dominated
by early type galaxies, it has not been clear up until now which
effect dominates; i.e., whether the increased clustering strength is
due to the larger luminosities of early types or to the dominance of
early types at bright luminosities. The recent results of
\citet{norberg:02} suggest the former; i.e. that the luminosity
dependence is the dominant effect. We therefore elect to use the
values from Table~3 of \citet{zehavi:02}, based on a volume-limited
low-redshift sample of very luminous ($-23.0 < M_{r^*}< -21.5$)
galaxies from the Sloan Digital Sky Survey (SDSS). While this sample
has not been explicitly selected by type or color, it is more likely
to be representative of the population in which we are interested than
the fainter, color selected sample presented in the same paper. The
values adopted (see Table~\ref{tab:data}) are similar to the results
for $L>L_*$ early-type/red galaxies in the literature
\citep[e.g.][]{loveday:95,hermit:96,willmer:98,brown:00,cabanac:00}. We
expect that a more secure determination of the clustering of luminous
{\em red} galaxies from the SDSS will be available shortly.

The situation for EROs is at least as problematic.  Unfortunately
there is no real consensus in the literature regarding what should be
deemed ``extremely red,'' and in any case any simple color cut is
bound to be somewhat arbitrary.  This confusion is compounded by the
variety of different filters used in these surveys.  We refer the
reader to \citet{firth:02} for transformations between several
commonly used sets of colors for typical galaxy spectra. Color cuts
commonly used to define EROs include $R-K>5$ or $6$, $R-H>4$ or $5$,
$I-H >3$ or $4$, $I-K>4$.  Another caution is that the surface
densities (and therefore number densities and spatial correlation
scales) all appear to be very sensitive to the precise color cuts and
magnitude limits assumed, and may be different for the ``old'' and
``dusty'' types.  The values we adopt are drawn from the measurements
of \citet{mcc:01} and \citet{firth:02}, which are based on galaxies
with $H < 20.5$ and $I-H>3$, corresponding approximately to a median
rest-frame B-band magnitude of $M_B = -20.3$ \citep{firth:02}.

Lyman Break galaxies (LBGs) are defined as in the sample of
\citet{adel:98}, using a color-color cut and an (AB) magnitude limit
of ${\mathcal R} < 25.5$ \citep{steidel:96b}.  We use the correlation
function parameters determined by \citet{adel:00proc}, and compute the
comoving number density using the selection function given by
\citet{steidel:99}; see BWS02 for details. We use the same overall
correction for incompleteness and the selection function adopted by
\citet{wechsler:01}.

The relevant clustering parameters for the dark matter, based on
equations \ref{eqn:dmr0} and \ref{eqn:dmg0}, are presented in
Table~\ref{tab:gifdata}, and the observables discussed above are
summarized in Table~\ref{tab:data}.

\begin{inlinefigure}
\begin{center}
\resizebox{\textwidth}{!}{\includegraphics{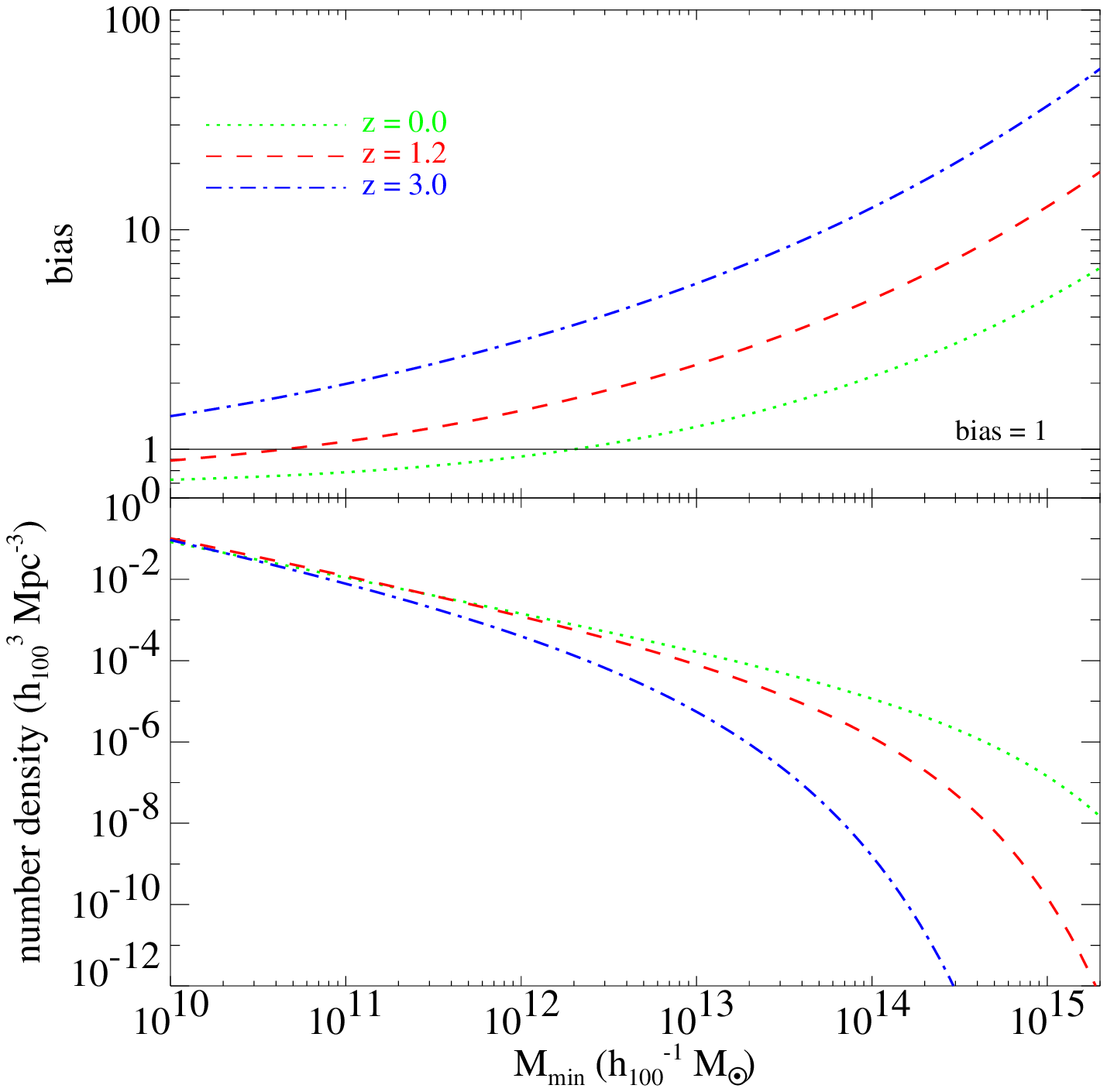}}
\end{center}
\figcaption{We plot the average bias and comoving number density as a
function of minimum host halo mass $\Mmin$, for the simple case of one
galaxy per dark matter halo above the minimum mass $\Mmin$.  This
corresponds to an occupation function with $\alpha=0$.
\label{fig:oneperhalo}}
\end{inlinefigure}

\section{An Analytic Model for Galaxy Clustering}\label{sec:models}

Our model has only two moving parts. We adopt analytic approximations
connecting the masses of dark matter halos with their number densities
and clustering strengths \citep{ps:74, mw:96, st:99}.  We then
parameterize the relationship between dark matter halos and their
galaxies (the occupation function) using a simple functional
form. This allows us to predict the number density and large scale
bias for a population with a given occupation function, or conversely,
to invert observed values of the number density and bias to obtain the
corresponding occupation function parameters.

For the halo mass function, we use the analytic expression developed
by \citet{st:99}, which agrees fairly well with the results of N-body
simulations \citep[see also][]{jenkins:01}:
\begin{equation}
\frac{{\rm d}n_h}{{\rm d}M} = - \frac{\bar{\rho}}{M} \frac{\rm{d} \sigma}{\rm{d} M}
\sqrt{\frac{a \nu^2}{c}} \left[ 1 + (a \nu^2)^{-p} \right] \exp\left[
\frac{-a \nu^2}{2}\right].
\end{equation} 
Here, $\sigma$ is the linear rms variance of the power spectrum on the
mass scale M at redshift $z$ and $\nu \equiv \delta_c / \sigma$, where
$\delta_c \simeq 1.686$ is the critical overdensity value for
collapse.  The other parameters are $a=0.707, p = 0.30,$ and $c=
0.163$, which were chosen to match N-body simulations with the same
cosmology and power spectrum as the one we have assumed. The comoving
number density of \emph{galaxies} is then given by the integral over
the halo mass function $dn_h/dM$, weighted by the appropriate galaxy
occupation function:
\begin{eqnarray}
\label{eqn:numdens}
n_{g} =  \int_{\Mmin}^{\infty} 
\frac{{\rm d}n_h}{{\rm d}M}(M)N_{g}(M) {\rm d}M.
\end{eqnarray}

We determine the large-scale bias for galaxies by integrating the
expected bias of halos as a function of mass $b_h(M)$, weighted by the
galaxy occupation function $\Ng$:
\begin{eqnarray} 
\label{eqn:bias}
b_g = 
\frac{1}{n_g} \int_{\Mmin}^{\infty} 
\frac{{\rm d}n_h}{{\rm d}M}(M) b_h(M) N_g(M) {\rm d}M. 
\end{eqnarray} 
For the halo bias $b_h$, we use the expression of \citet{st:99}:
\begin{equation}
\label{eqn:halo_bias}
b_h(M) = 1 + \frac{a \nu^2 - 1}{\delta_c} + \frac{2 p / \delta_c}{1 +
(a \nu^2)^p}.
\label{eqt:stb}
\end{equation}

The final missing piece is the galaxy occupation function $N_g(M)$, or
the number of observed galaxies per halo, at a given magnitude limit
and redshift, and as a function of halo mass.  
In this paper, we choose a simple power law form, with a
normalization $\Mn$, a slope $\alpha$ and a low-mass cutoff $\Mmin$:
\begin{equation}
N_g(M>\Mmin) = \left(\frac{M}{M_1}\right)^{\alpha}.\label{eqn:occfn}
\end{equation}
This functional form is a reasonably good approximation to the
occupation function predicted by semi-analytic and hydrodynamic models
(\citealt{wechsler:01}; \citealt*{whitem:01}).  Physically, the
parameter $\Mmin$ represents the smallest mass halo that can
\emph{ever} host a galaxy, and $\Mn$ represents the mass of a halo
that will host on average one galaxy. The slope $\alpha$ represents
how strongly the number of galaxies per halo depends on halo mass. As
one expects larger mass halos to host more galaxies than smaller mass
halos, we restrict our analysis to $\alpha \geq 0$. The large-scale
linear bias $b_g$ is independent of the normalization of the
occupation function $M_1$ and determines $\Mmin$ (for a fixed value of
$\alpha$; see Fig.~\ref{fig:tables}). The normalization $M_1$ is then
fixed by the required number density.  We also define the
(galaxy-number weighted) ``average'' host halo mass, $\Mbar$:
\begin{equation}
\Mbar =
\frac{1}{n_g} \int_{M_{min}}^{\infty}
M\, \frac{dN_h}{dM}(M,z) N_g(M, z) dM. 
\end{equation}

Once we have determined the occupation function parameters using a set
of observations at a fixed redshift, we can then calculate the
observables (correlation length and number density) for populations at
different redshifts, selected either assuming a constant minimum mass
or constant number density (\S\ref{sec:invert}). This can be used to
determine whether observed populations at different redshifts are
likely to occupy halos with similar masses --- for example, to
determine if the halos occupied by EROs are similar in mass to those
that host nearby giant ellipticals. This is discussed in
\S\ref{sec:zev}.

\section{Solving for the Galaxy Occupation Function Parameters}
\label{sec:invert} 

In this section, we use the observational constraints outlined in
\S\ref{sec:data} and the formalism summarized in \S\ref{sec:models} to
obtain constraints on the allowed range of values for the occupation
function parameters $\Mmin$ and $\Mn$. We carry out the analysis at
redshift $z=0$ for giant ellipticals, at $z=1.2$ for EROs, and at
$z=3$ for LBGs. Subsequently, we will investigate the implications for
interpreting the connections between these populations.

First, in Fig.~\ref{fig:oneperhalo} we show the average bias and
number density for the simple case of one galaxy per halo, for halos
above a minimum mass $\Mmin$, as a function of $\Mmin$ (from
Eqns.~\ref{eqn:numdens} and \ref{eqn:bias}, with $N_g=1$). This
corresponds to the case of $\alpha=0$ in our model, in which case the
parameter $\Mn$ becomes undefined, and the clustering and number
density are solely determined by $\Mmin$. Note that for a given value
of the bias, the \emph{maximum} value of $\Mmin$ is obtained for this
one-galaxy-per-halo ($\alpha=0$) case. From this, we may read off
upper limits on $\Mmin$ for $z=0$ gEs ($1.7\times10^{13} \hmsun$),
$z=1.2$ EROs ($1.5 \times 10^{13} \hmsun$), and $z=3$ LBGs ($7.7\times
10^{10}\hmsun$). These values are summarized in
Table~\ref{tab:masses}. It is interesting to note that the observed
correlation length of present-day rich X-ray clusters is $r_0 \simeq
20$ \hmpc\ \citep[e.g.][]{collins:00}, implying a very large bias of
$b \sim 10$. From Fig.~\ref{fig:oneperhalo}, we would infer masses of
$\Mmin \gsim {\rm few} \times 10^{15} \hmsun$ for this population,
consistent with X-ray and lensing masses.

\begin{inlinefigure}
\begin{center}
\resizebox{\textwidth}{!}{\includegraphics{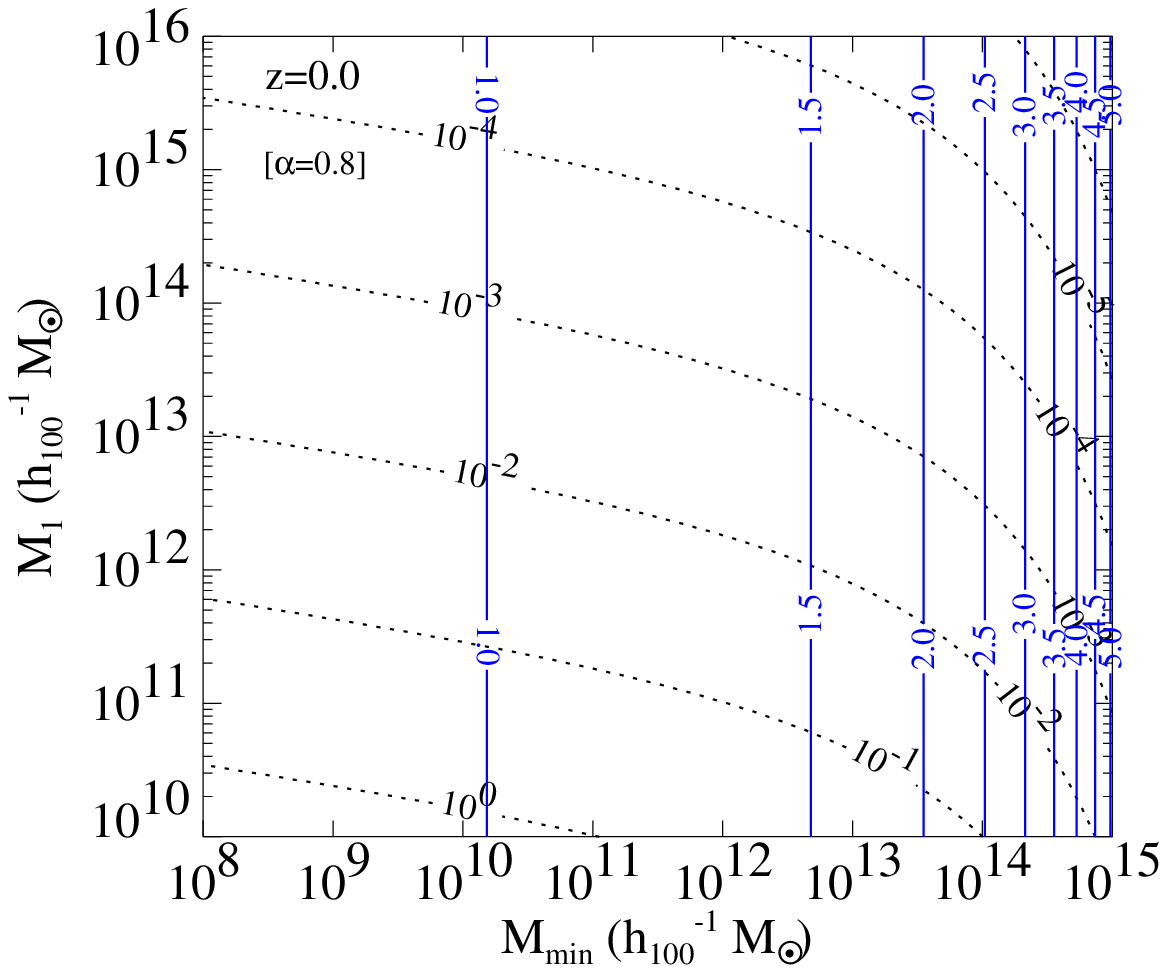}}
\resizebox{\textwidth}{!}{\includegraphics{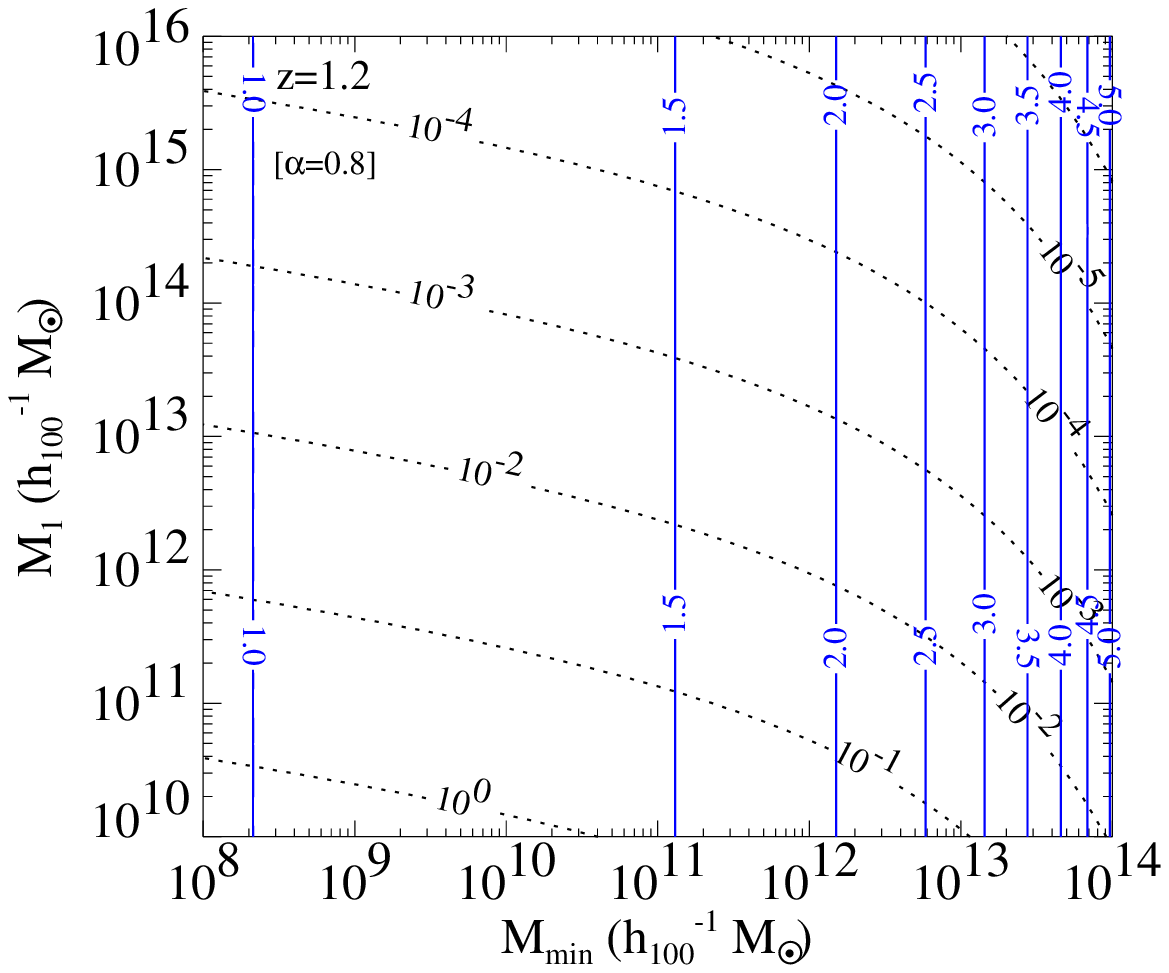}}
\resizebox{\textwidth}{!}{\includegraphics{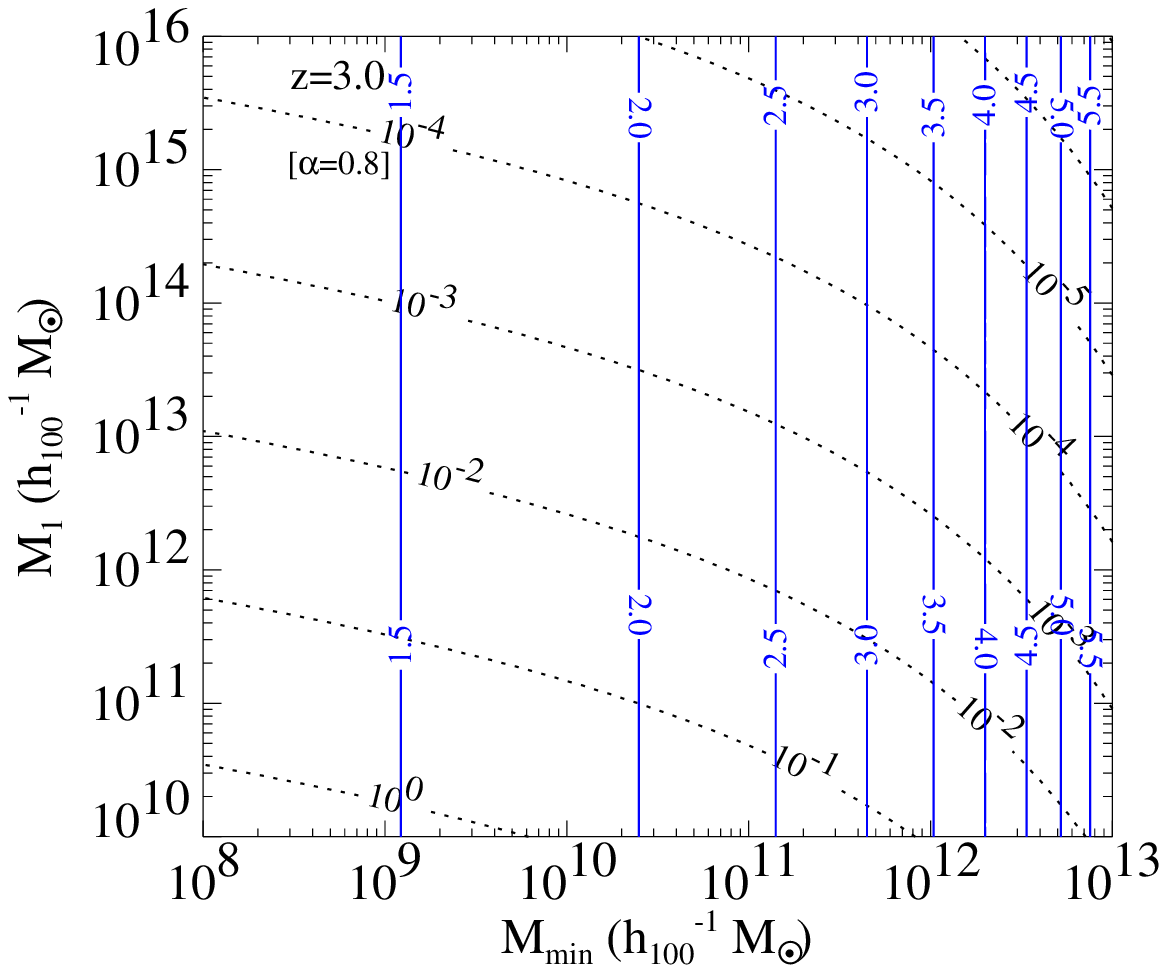}}
\figcaption{The occupation function parameters for a fixed assumed
slope, $\alpha=0.8$, at redshifts $z=0.0, 1.2, 3.0$ (top, middle, and
bottom panels).  The constant-bias lines (vertical {\bf solid} lines)
are independent of $\Mn$ (which represents the mass of a halo that
will host on average one galaxy).  Combined with a measurement of the
number density (curved {\bf dotted} lines, in units of
h$_{100}^{3}$\,Mpc$^{-3}$) of a galaxy population, both $\Mn$ and
$\Mmin$ are uniquely determined (for an assumed slope $\alpha$).
\label{fig:tables}}
\end{center}
\end{inlinefigure}

In Fig.~\ref{fig:tables}, we show the solutions in the two-dimensional
parameter space $\Mmin$ vs. $\Mn$ for fixed values of the observed
galaxy number density $n_g$ and bias $b_g$, at the redshifts of
interest for our three populations. For purposes of illustration, we
have fixed the slope of the occupation function to $\alpha=0.8$.
Whereas in the future it may be possible to use spatial clustering
properties on small scales to constrain the value of $\alpha$ (see
BWS02), in this paper we must treat $\alpha$ effectively as a free
parameter. We note, however, that in semi-analytic models
\citep{sp:99, kauf:99b}, the value of $\alpha$ is close to $0.7-0.9$
for bright galaxies, and so will concentrate on a fiducial value
$\alpha=0.8$ in parts of the discussion. Fig.~\ref{fig:tables} allows
us to read off the values of $\Mmin$ and $\Mn$ for any desired values
of the observables $n_g$ and $b_g$. For example, at $z=0$, using the
values of $n_g$ and $b_g$ for $L>L_*$ ellipticals from
Table~\ref{tab:data}, we can see these galaxies occupy halos with
masses greater than $\Mmin \simeq$\,few\,$\times 10^{13} \hmsun$
(Table~\ref{tab:masses}). This is consistent with the predictions of
semi-analytic models \citep{somerville:01, benson:00}, and with
observationally determined group/cluster masses.

\begin{figure*}
\begin{center}
\resizebox{0.47\textwidth}{!}{\includegraphics{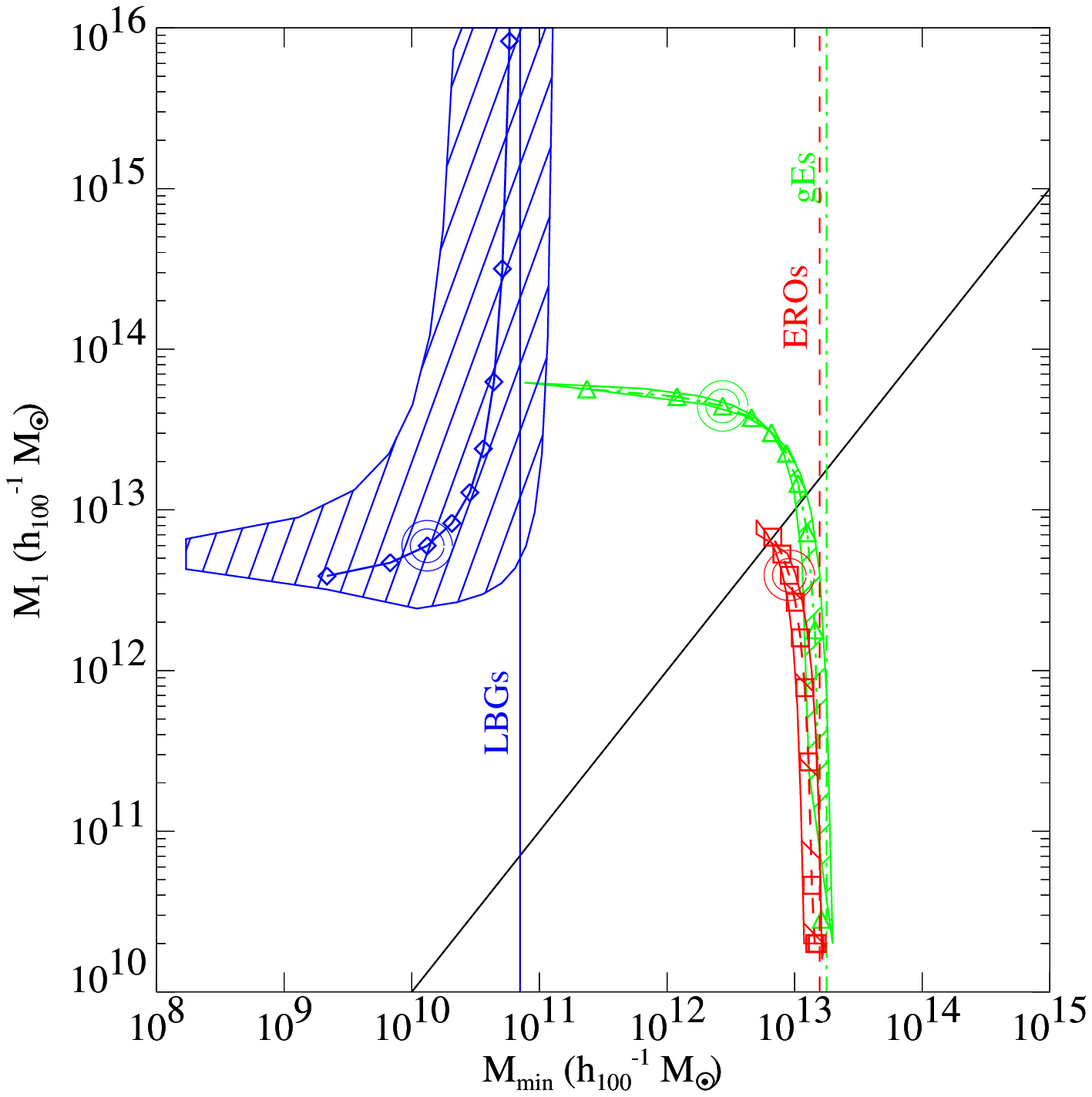}}
\resizebox{0.47\textwidth}{!}{\includegraphics{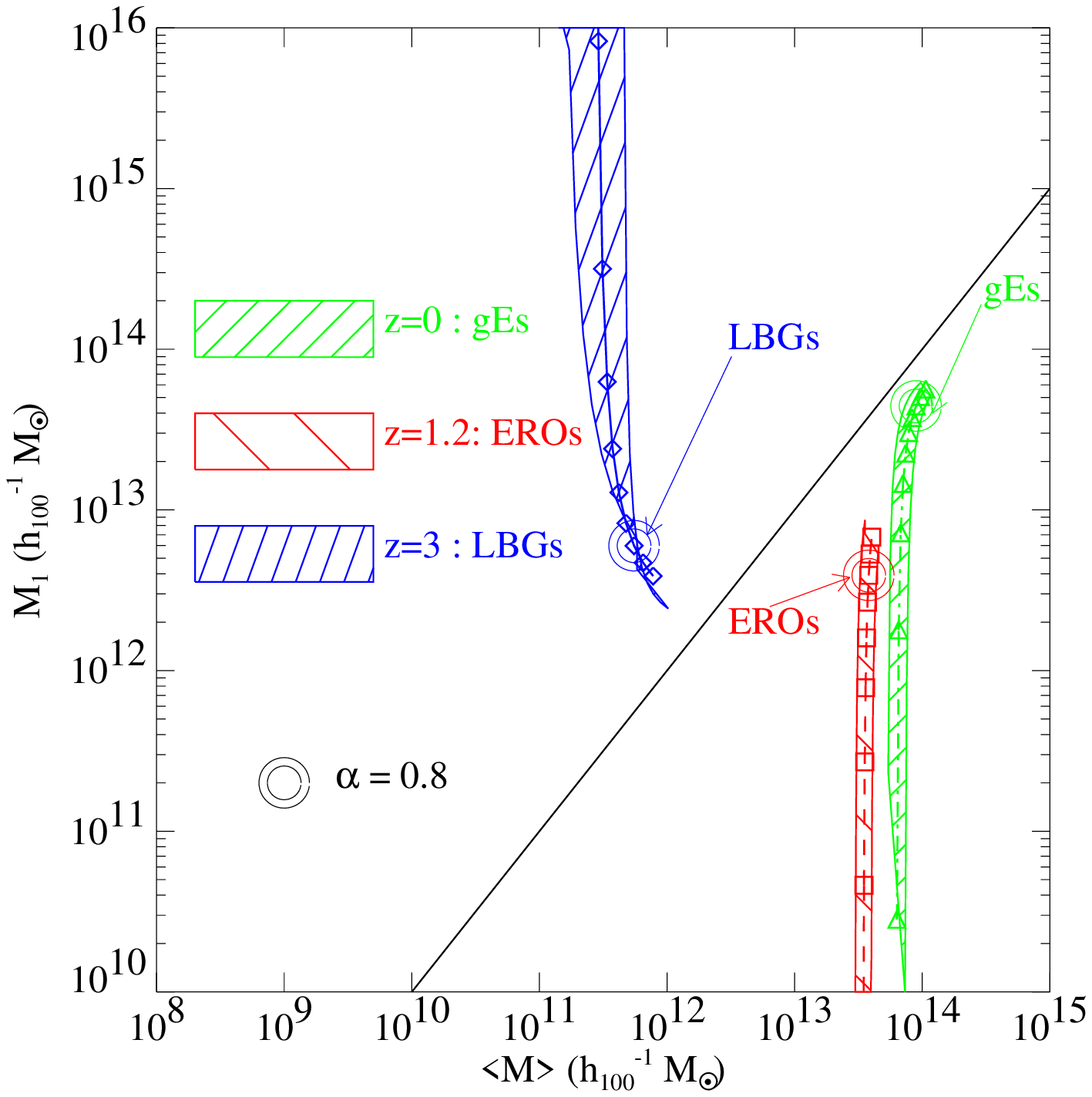}}
\caption{The explicit inversion of the clustering and number density
parameters for each galaxy population at its redshift recovers the
minimum halo mass and the halo-mass weighted mass, $\Mmin$ ({\bf left
panel}) and $\Mbar$ ({\bf right panel}) respectively. This calculation
was done for the full range of uncertainties in all observational
quantities, as discussed in \S\ref{sec:data} and summarized in
Table~\ref{tab:data}.  Each point indicated corresponds to a different
value for the slope of the occupation function, with the range
$\alpha=0-1.0$ in intervals of $0.1$.  A value of $\alpha\approx0.8$
(shown by the double-circle points) is shown.  The {\bf dot-dashed}
lines with {\bf triangles} are for $z=0$, $L>L_*$ Elliptical galaxies
(gEs); the {\bf dashed} lines with {\bf squares} are for $z\approx1.2$
EROs; and the {\bf solid} lines with {\bf diamonds} are for
$z\approx3$ LBGs.  The labelled vertical lines mark the $\alpha=0$
values for $\Mmin$ (see Table~\ref{tab:masses}). Note that the $\Mmin$
and $\Mbar$ results are relatively insensitive to the occupation
function slope $\alpha$ for the strongly-clustered EROs and luminous
ellipticals.  The solid diagonal lines correspond to $\Mmin=\Mn$ and
$\Mbar=\Mn$.
\label{fig:invert}}
\end{center}
\end{figure*}

In Fig.~\ref{fig:invert}, we show the solutions for the occupation
function parameters for the three populations, for different values of
the slope $\alpha$. Higher values of $\Mmin$ correspond to lower
values of $\alpha$. As noted above, the maximum value for $\Mmin$ is
obtained for the limiting assumption of one object per halo,
$\alpha=0$, and is depicted by labelled vertical lines for each case,
for the central values of Table~\ref{tab:data}.  The filled areas show
the full range of parameter space allowed by the inversion of the
chosen values for the number density and correlation length of each
population with their stated ($1-\sigma$) uncertainties
(Table~\ref{tab:data}).  It should be noted that we have simply used
the uncertainties from the literature {\em for a given sample}, and
have not attempted to include the uncertainty/spread due to
sample-to-sample variance or differences in sample selection. These
effects will actually lead to a much larger region of allowed
parameter space.  The minimum mass $\Mmin$ is almost independent of
$\alpha$ for the strongly clustered local ellipticals and EROs. For
the more weakly clustered LBGs, the minimum halo mass is a stronger
function of $\alpha$. We also show the average host halo mass $\Mbar$
(weighted by galaxy-number), which is even less dependent on $\alpha$
for all cases. From this figure, we see that the clustering and number
density of EROs are consistent with their occupying halos of similar
masses to those that harbor present-day $L>L_*$ ellipticals. In
contrast, LBGs must occupy halos that are several orders of magnitude
smaller in mass.

The relationship of the values of $\Mmin$ and $\Mbar$ to those of
$\Mn$ is also interesting to consider. Recall that $\Mn$ corresponds
to the mass of a halo that harbors on average one galaxy --- therefore
halos with masses greater than $\Mn$ will tend to contain more than
one galaxy.  The case $\Mn >\!\!> \Mmin$ (above the diagonal lines in
Fig.~\ref{fig:invert}) corresponds to a situation where most galaxies
are the sole occupants of their dark matter halos. The LBGs fall into
this category. Conversely, for $\Mn \lsim \Mmin$, most galaxies have
one or more companions within a common dark matter halo. The EROs and
gEs are in this category. This is interesting, as it provides indirect
evidence for the general concept of hierarchical clustering (i.e.,
that large structures like groups and clusters are built up over time
through merging), and also is even more directly suggestive that EROs,
like gEs, are found in group/cluster type environments and may have
been influenced by interactions with other galaxies.

Although LBGs are not the main focus of this paper, as they have been
the topic of so many analyses of this type, and because our results
differ significantly from some of those in the literature, it is worth
a brief digression to discuss the reasons for this difference. Some
past analyses have used the clustering of LBGs to argue that they are
harbored by halos with considerably larger masses of $\simeq
10^{12}\hmsun$ \citep[e.g.][]{steidel:96b, giav:98}.  One reason for
this is that early results on the observational clustering values
based on a few fields yielded considerably larger correlation lengths,
$r_0=6$\hmpc\ \citep{adel:98}, corresponding to bias values of
$b\approx2.7$ (using our definition of bias). Apparently, the LBGs in
these fields showed an unusually strong degree of clustering, and more
recent results based on considerably larger areas have yielded lower
values; for example, \citet{gd:01} find $r_0 = 3.2\pm0.7$\hmpc, and
\citet{adel:00proc} finds $r_0 = 3.8\pm0.3$\hmpc\ \citep[see the
broader compilation and discussion in][]{wechsler:01}. The
corresponding values of the linear bias (using our standard
definition) are $b \sim 1.4$--1.9.  Differing definitions of bias lead
to a still broader range of values; for example \citet{adel:98} adopt
$b \sim 4$ for the same \lcdm\ model.  In addition, most previous
analyses of this kind have made the simplifying assumption that each
halo above a given mass threshold $\Mmin$ contains exactly one galaxy,
corresponding to the limiting $\alpha=0$ case discussed above. As one
can see from Fig.~\ref{fig:oneperhalo}, had we assumed $b \sim 4$ and
$\alpha=0$ for LBGs as was done in \citet{adel:98} we would have found
$\Mmin \sim 10^{12}\hmsun$, consistent with the value obtained by
those authors. Note that our results are also consistent with those of
\citet{wechsler:01}, in which a detailed treatment of LBG clustering
based on semi-analytic modeling combined with N-body simulations was
carried out, and with BWS02, which used an analytic model similar to
the one used here. BWS02 obtained further constraints on the slope of
the occupation function $0.9\lsim \alpha \lsim 1.1$ from small scale
clustering data\footnote{Although BWS02 use the same values as we do
for the observed correlation function parameters of LBGs, they use a
different definition of bias. They chose to assume that the slope of
the dark matter correlation function is the same as that of the galaxy
correlation function, rather than defining the bias at a fixed scale,
as we have done. This is why the fiducial value of the bias for LBGs
in BWS02 is 2.4 while we adopt $b_{\rm LBG} = 1.9$. }.

\section{Relating Populations at Different Redshifts}
\label{sec:zev}
It is always tempting to speculate on the connection of a population
observed at high redshift with the more familiar objects in the local
Universe. This is a complicated and somewhat ambiguous task in the
context of the hierarchical scenario, in which objects do not keep
their identity, but are constantly merging and can even change their
Hubble type. We cannot try to ``follow'' the properties of a
particular object through time with the simple models presented here,
but we can place some constraints on the connections of the three
populations we have been discussing with one another, and we can ask
specific questions about populations selected in certain ways at
different redshifts. This is the topic of this section. We consider
two types of models: a classical ``galaxy conserving'' model, and
hierarchical merging models for objects selected with a fixed mass
threshold, or with a fixed number density.

\begin{inlinefigure}
\begin{center}
\resizebox{\textwidth}{!}{\includegraphics{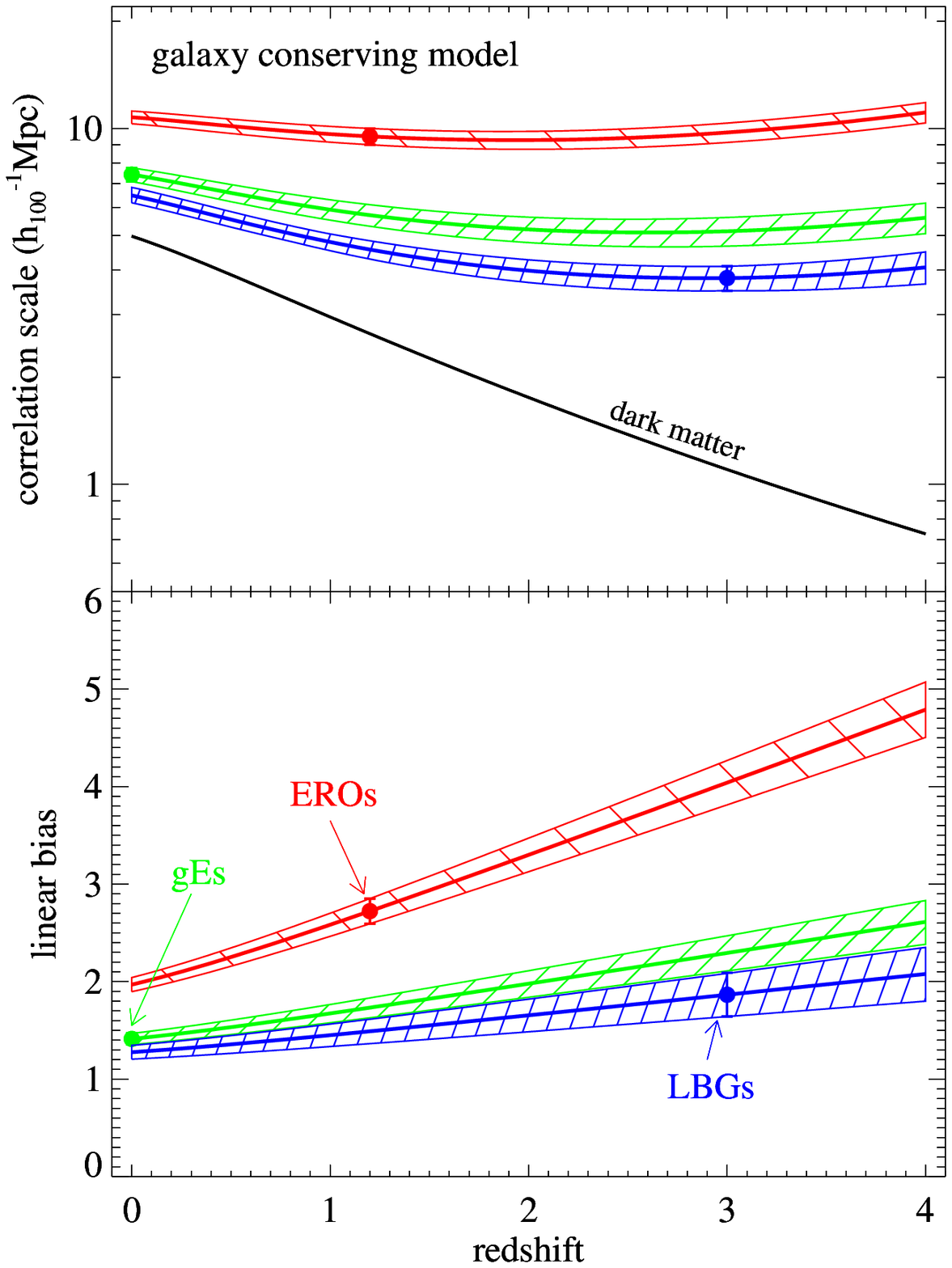}}
\end{center}
\figcaption{The redshift evolution of the comoving correlation scale 
$r_0$ (top); and the linear bias (lower panel), in the \emph{galaxy
conserving model}, for populations as indicated.  The filled regions
correspond to measured uncertainties, as described in the text and in
Fig.~\ref{fig:zevol}.  The dark {\bf solid} line in the upper panel
shows the correlation scale of the dark matter.
\label{fig:gcevol}}
\end{inlinefigure}

\subsection{The galaxy conserving model
\label{subsec:galconsmod}}
First, we consider a very simple null-hypothesis model for describing
clustering evolution.  If galaxies are assumed to form with some
``bias at birth'' and then to evolve without merging or changing their
luminosity, their relative positions will change solely as dictated by
gravity.  This is a fluid flow problem, and the behavior of galaxies
as tracers of the underlying matter will obey the continuity and Euler
equations \citep{peebles:80}.  \citet{fry:96} worked out this
conceptually simple ``galaxy conserving'' model, wherein the linear
bias evolves by
\begin{equation}
b(z) = 1 + (b_0 - 1)/D(z).
\end{equation}
Here, $D$ is the growth factor of the universe for our adopted
cosmology, and $b_0$ is the ``bias at birth'' of the population.  For
$b_0$ we can use the bias values at the redshift of observation, and
then trace the evolution forward and backward in time again.  Implicit
in this model is that there is no merging or any loss of identity for
any of the galaxy-particles \citep[see also][]{tegmark:98}.  In this
model, as the universe expands, the bias decreases monotonically and
the correlation scale tends towards that of the underlying dark matter
\citep{fry:96}.  The results using this model are shown in
Fig.~\ref{fig:gcevol}. If LBGs evolved in this way, they would end up
being slightly less clustered than $z=0$ gEs. The extrapolation of
EROs' clustering properties to $z\approx0$ suggest that they would
correspond to a \emph{very} strongly clustered local population, even
more biased than $L>L_*$ ellipticals today.

\begin{figure*}
\begin{center}
\resizebox{0.47\textwidth}{!}{\includegraphics{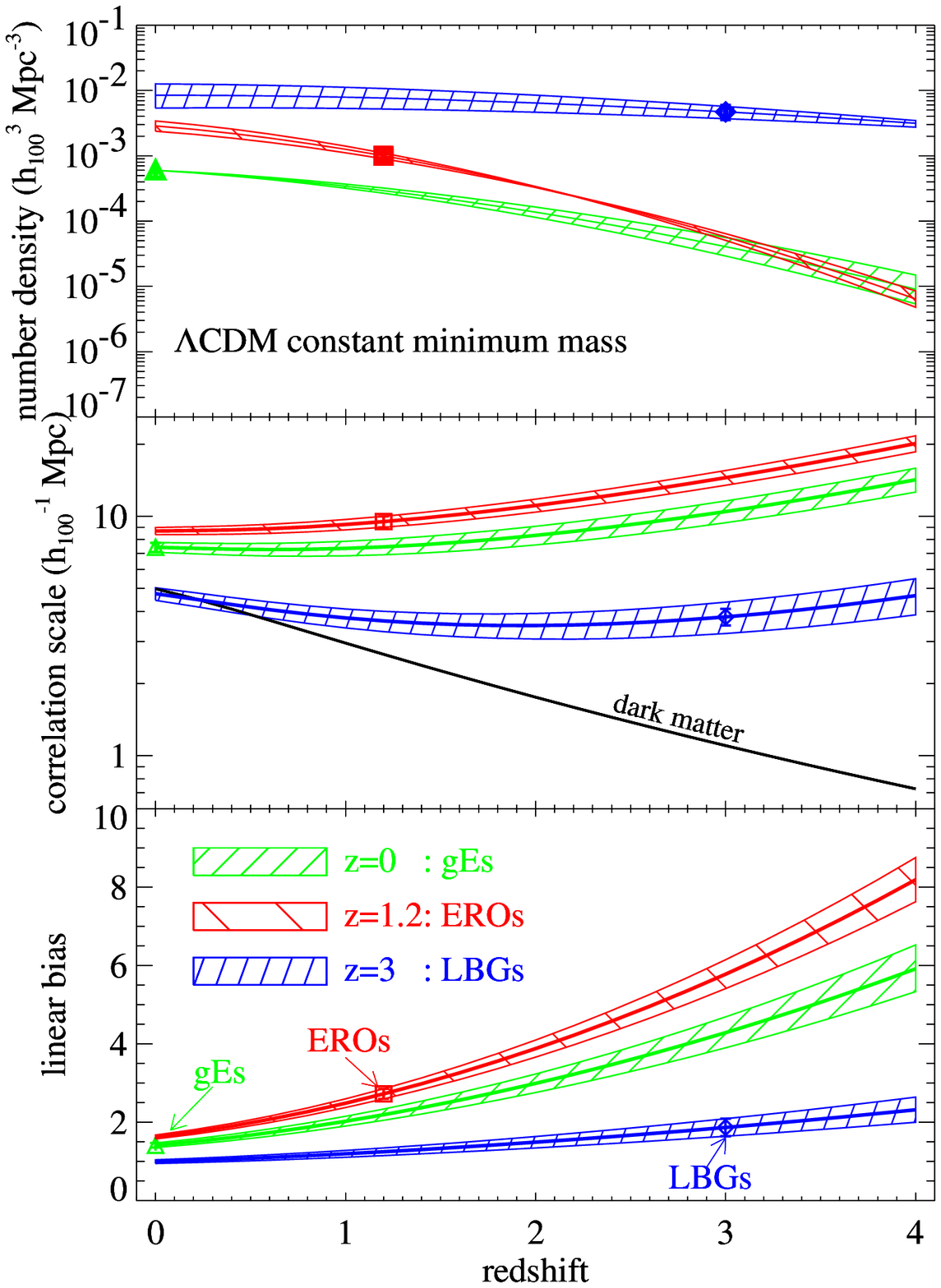}}
\resizebox{0.47\textwidth}{!}{\includegraphics{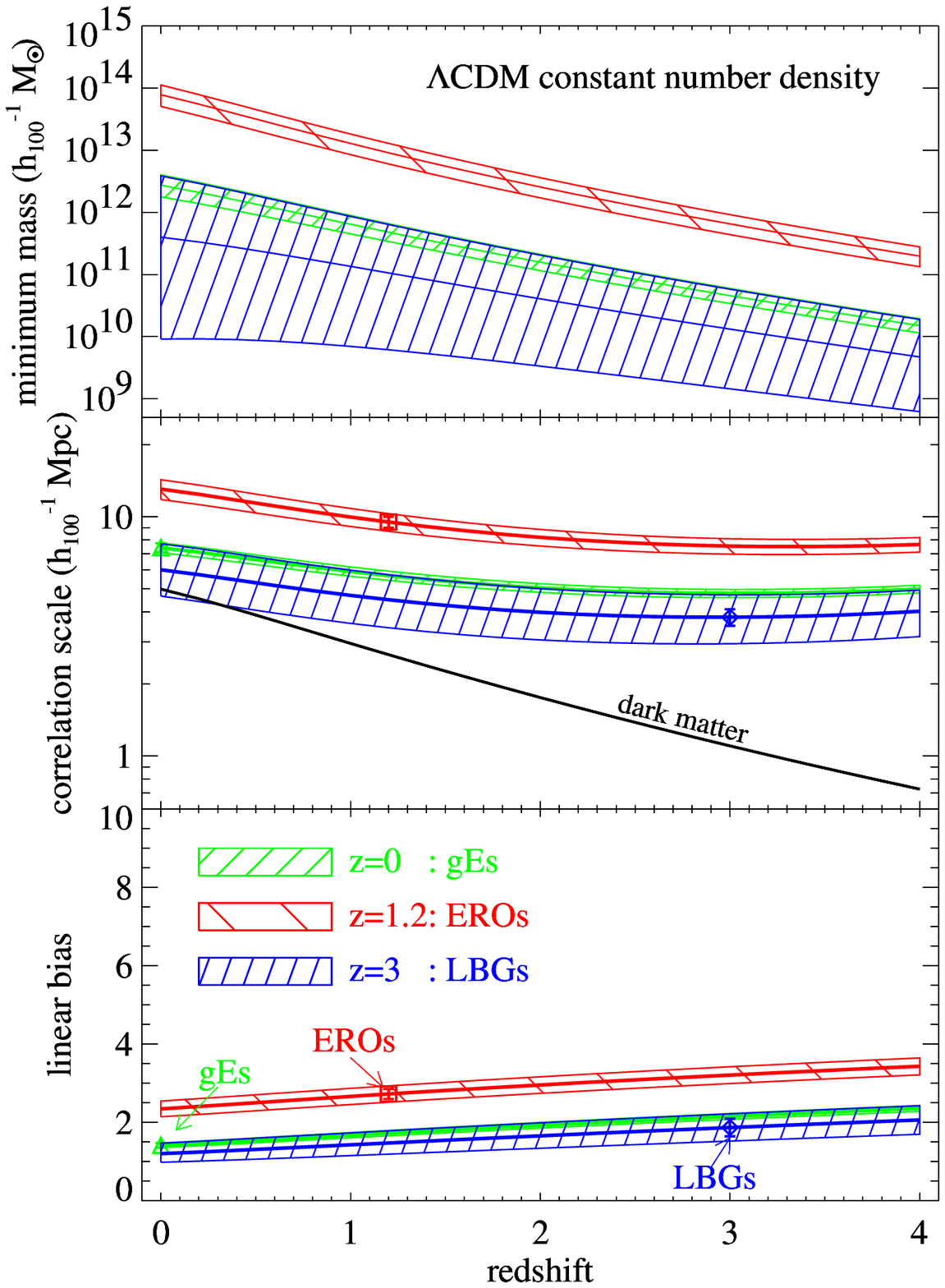}}
\caption{{\bf left:} In the {\em constant minimum mass} model, the
redshift evolution of the comoving number density (top panel); the
comoving correlation scale $r_0$ (middle panel); and the linear bias
(lower panel).  Each filled region reflects the uncertainties in the
values of the occupation function parameters (as shown graphically in
the inversion figures of Fig.~\ref{fig:invert}), propagated with
redshift by the recipe presented in the text.  {\bf right:} In the
{\em constant number density} model, the top panel shows the redshift
evolution of $\Mmin$; the second and third panels are as in the left.
The dark matter correlation scale (Eqn.~\ref{eqn:dmr0}) is shown as a
solid dark line.  These plots are computed for $\alpha=0.8$.
\label{fig:zevol}}
\end{center}
\end{figure*}

\subsection{Hierarchical Merging models
\label{subsec:mergmod}}

We now use the hierarchical framework presented earlier to produce
predictions for the clustering properties of populations selected at
different redshifts in two different ways. In the first case, we
select objects with a fixed mass threshold ({\bf{\em constant minimum
mass}}, ``CMM''). Here, we assume that all parameters of the
occupation function remain the same, and that halos above a fixed mass
threshold ($\Mmin$) may host galaxies, according to
Eqn.~\ref{eqn:occfn}. We use the occupation function parameters
obtained in \S\ref{sec:invert} for each of the three populations, and
run these values forward and backward in time to predict clustering
properties over a range of cosmological epochs. We also consider the
properties of populations selected to have a fixed number density at
all redshifts ({\bf{\em constant number density}}, ``CND''). Here, we
assume that the slope and normalization of the occupation function
remain constant over time, but that the minimum mass required to host
a galaxy changes, such that the comoving number density of galaxies
remains constant. This can be directly connected with observations.
As discussed at the beginning of this section, we note again that
neither of these models is meant to represent the actual evolution in
time of a particular object. Instead, we represent the evolution of
statistically defined samples, which may be related to similarly
defined observed samples.

Fig.~\ref{fig:zevol} shows how the number density or minimum mass,
correlation length, and bias change with time for populations selected
in these two ways. We have kept the occupation function slope fixed to
$\alpha=0.8$, and we have assumed that the slope of the galaxy
correlation function $\gamma_g$ is constant with redshift. In the
constant mass case, because halos of a fixed mass represent rarer
fluctuations in the density field at earlier times, the bias decreases
rapidly with time. In the constant number density case, the required
$\Mmin$ {\em increases} with time, so that the bias evolves much less
rapidly, but still in the same sense. 

In the CMM case, the mass-selected ERO-like population run forward in
time are predicted to be slightly more numerous and more clustered
than the $z=0$ gEs. This apparently contradictory result is due to
the fact that we found that for $\alpha=0.8$, EROs occupy slightly
more massive halos than gEs, but $\Mn$ was considerably larger
because of the higher number density we adopted for EROs. Concurrent
evolution in $\alpha$, $\Mn$, and/or $\gamma_g$ could change the
details of this prediction; and moreover the value of the number
density is quite uncertain. Generally though, this result supports the
idea that EROs {\em could} be the already mostly assembled progenitors
of $z=0$ luminous giant ellipticals. Conversely, the mass-selected
LBG-like population is much more numerous and less clustered than
either EROs or gEs when run forward in time to $z=0$, because of the
relatively low-mass halos they occupy in our model. These results
suggest that if LBGs are indeed the progenitors of gEs, they must
undergo significant mass-growth through merging between $z=3$ and the
present. More speculatively, this perhaps suggests that early type
galaxies may have been mostly assembled by $z\sim1$, but were in much
smaller pieces at $z\sim3$ (or else their progenitors are not
associated with the LBGs).

In the CND case, the number-density selected ERO-like objects
projected forward in time end up in much more massive halos, and
therefore much more clustered, than $z=0$ gEs (but are much less
massive and clustered than rich X-ray clusters). The masses and
clustering properties of the number-density selected LBG-like objects
projected to $z=0$ overlap those of the gEs. This may lead to the
same conclusion as the above: that LBGs may evolve into gEs by
$z=0$, but must grow in mass considerably.

\section{Summary and Conclusions}\label{sec:conclusions}
Extremely Red Objects are an intriguing population which may be the
$z\approx1$ progenitors of giant ellipticals.  Recent results from
Wide-Field Near-IR selected surveys with multi-band photometry have
made it possible to obtain angular clustering and photometric redshift
estimates for these objects, allowing the number density and
real-space correlation length to be derived from the angular
clustering and surface density \citep{daddi:01,mcc:01,firth:02}. These
data provide important clues about the nature of EROs by constraining
the masses of halos that harbor them in a \lcdm\ hierarchical model of
structure formation.  We have used an analytic model, combining a
simple parameterization of the occupation function of dark matter
halos with analytic approximations for halo number density and bias,
to constrain the masses of halos harboring EROs. We apply the same
approach to $z=0$ ellipticals and $z=3$ LBGs, and using this
information, speculate on the connection between these populations and
EROs. Our main conclusions may be summarized as follows:
\begin{itemize}
\item EROs at $z\sim1.2$ occupy halos with minimum masses $\Mmin
\simeq 10^{13}\hmsun$, with a galaxy-number weighted average halo mass
of $\Mbar\simeq 5 \times 10^{13}\hmsun$ --- in other words, EROs are
probably located in rich groups or clusters, at $z\approx1$.  These
masses are lower limits for the subset of ``old'' EROs, if they are
associated with larger correlation lengths.  These results are rather
insensitive to the assumed value of the slope of the occupation
function $\alpha$.

\item Elliptical galaxies at $z\sim0$ with $L>L_*$ occupy halos with
very similar masses to those harboring EROs, lending support to the
hypothesis that EROs may be the {\em already almost fully assembled}
progenitors of present day giant ellipticals.

\item Recent results on the measured clustering of Lyman-break
galaxies at $z\sim3$ imply that they occupy much smaller mass halos
than EROs or local ellipticals, with $\Mmin \approx 10^{10-11}\hmsun$
and $\Mbar\approx 10^{11-12}\hmsun$. This does not necessarily imply
that LBGs are \emph{not} the progenitors of present-day $L>L_*$
ellipticals, only that their halos must experience considerable mass
growth via merging if indeed LBGs are to grow into giant ellipticals.
\end{itemize}

\acknowledgments 

LAM would like to thank the Institute of Astronomy, Cambridge, UK and
the University of Michigan for their hospitality during the course of
this work.  LAM was supported by the PPARC Rolling Grant
PPA/G/O/1999/00193 at the University of Oxford.  RSS was supported by
a PPARC theory rolling grant at the IoA. We thank Risa Wechsler, James
Bullock, Ofer Lahav and Andrew Firth for useful discussions. We thank
the referee, David Weinberg, for useful comments and suggestions that
significantly improved this paper.

\clearpage
% redefine the command that creates the equation number
\renewcommand{\theequation}{A\arabic{equation}}
\setcounter{equation}{0}  % reset counter 
\renewcommand{\thefigure}{A\arabic{figure}}
\setcounter{figure}{0}  % reset counter 
\section*{APPENDIX}  % use *-form to suppress numbering

The correlation function of the dark matter in our chosen cosmology
has been computed directly from the GIF/VIRGO N-body \lcdm\ simulation
\citep{jenkins:98}, and is shown in Fig.~\ref{appfig:xidm} for several
redshifts.  Approximating the correlation function on linear scales
with a power-law, we parametrize the comoving correlation length scale
$r_0$ and the correlation function slope $\gamma$ as a function of
redshift in a convenient analytic form:
\begin{eqnarray}
\log_{10}({\rm h}_{100}^{-1}\,r_{0_{\rm DM}}) & 
 = & \log_{10}(5.00) - 0.00142 - \nonumber \\ 
 & & 0.419\,\log_{10}(1+z) - \nonumber \\
 & & 1.11\,\log_{10}(1+z)^{2},\label{eqn:dmr0}\\
\gamma_{0_{\rm DM}}  & = & -1.74 + 0.169\,z.\label{eqn:dmg0}
\end{eqnarray}

\begin{inlinefigure}
\begin{center}
\resizebox{0.47\textwidth}{!}{\includegraphics{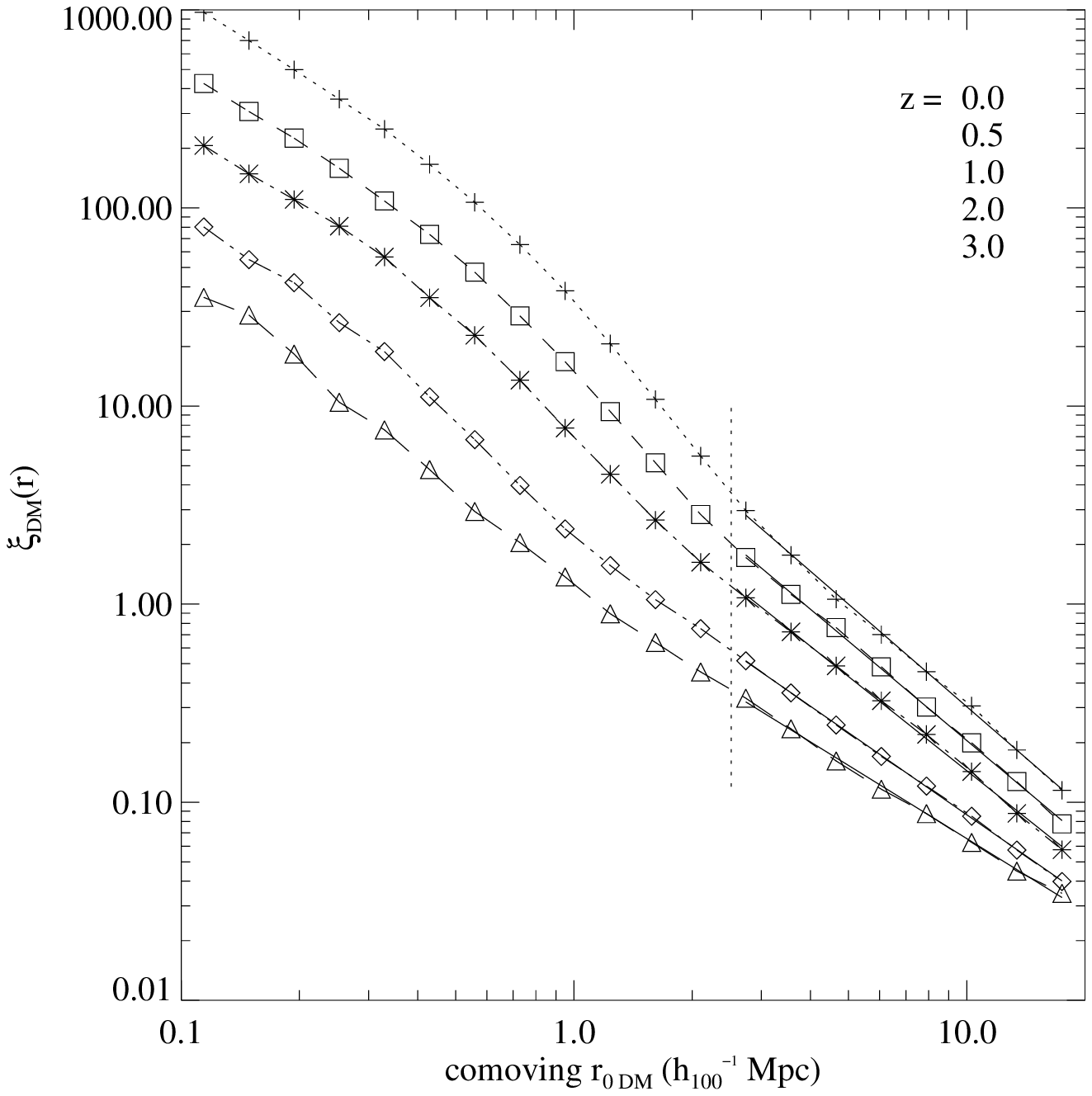}}
\resizebox{0.47\textwidth}{!}{\includegraphics{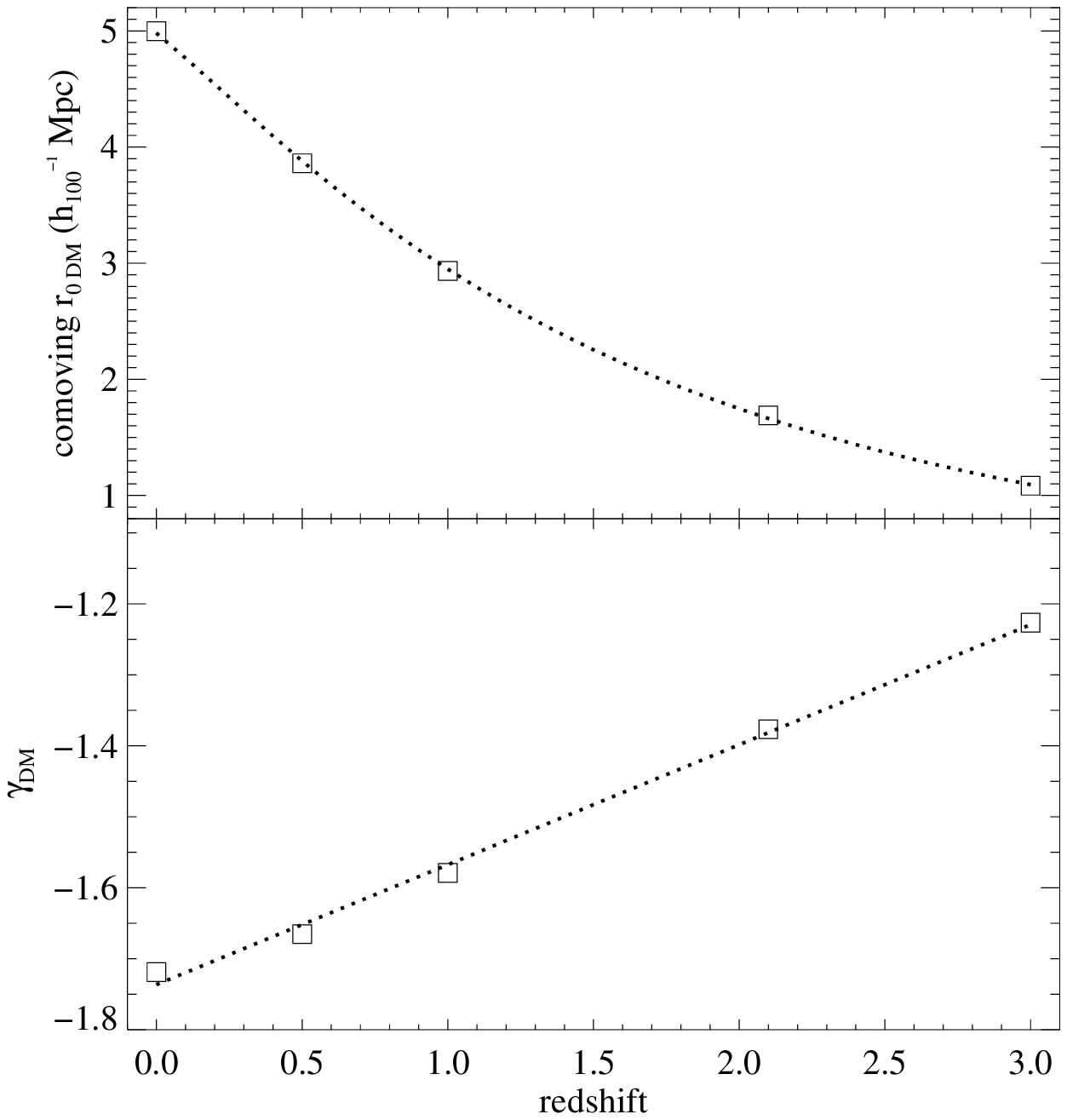}}
\end{center}
\figcaption{{\bf left:} The spatial correlation function, $\xidm$, of
dark matter based on the GIF/VIRGO \lcdm\ simulation, computed
directly at redshifts 0.0, 0.5, 1.0, 2.0, and 3.0.  The correlation
strength decreases with increasing redshift, as shown in the legend.
Linear regressions are computed for $r_{\rm
0,DM}>2.5$\,h$_{100}^{-1}$\,Mpc (to the right of the dotted line), and
are shown as solid lines over the measurements.  From these fits,
relevant for the $r=8$\,h$_{100}^{-1}$\,Mpc scale considered in this
paper, the correlation length ($r_{\rm 0,DM}$) and power ($\gamma_{\rm
DM}$) are computed.  Best-fitting polynomial fits are given in
Eqns.~\ref{eqn:dmr0} and \ref{eqn:dmg0}, and shown with the data in
the {\bf right} panel.
\label{appfig:xidm}}
\end{inlinefigure}

\clearpage

\begin{center}
\begin{table*}
\begin{center}
\caption{\label{tab:gifdata}Dark Matter Clustering Properties}
\begin{tabular}{rccc}%\hline
Parameter & $z=0$ & $z=1.2$ & $z=3$\\
\hline
\hline
$\xi_{8,{\rm DM}}$ & 
  $\phantom{-}0.439$ & 
  $\phantom{-}0.173$ & 
  $\phantom{-}0.087$ \\
$r_{0,{\rm DM}} \,({\rm h}_{100}^{-1}\,{\rm Mpc})$  & 
  $\phantom{-}4.981$ & 
  $\phantom{-}2.513$ & 
  $\phantom{-}1.101$ \\
$\gamma_{\rm DM}$ & 
  $-1.736$& 
  $-1.517$& 
  $-1.229$\\
\hline
\end{tabular}
\tablecomments{The correlation amplitude at 8\,h$_{100}^{-1}$\,Mpc
($\xi_{8,{\rm DM}}$), correlation length ($r_{0,{\rm DM}}$), and
correlation function slope ($\gamma_{\rm DM}$), for the dark matter,
at the redshifts of interest, as calculated from Eqns~\ref{eqn:dmr0}
and \ref{eqn:dmg0}, based on the GIF $\Lambda$CDM simulation
\citep{jenkins:98}.}
\end{center}
\end{table*}
\end{center}

\begin{center}
\begin{table*}
\begin{center}
\caption{\label{tab:data}Observational Values used for Analysis}
\begin{tabular}{cccc}%\hline
Parameter & gEs & EROs & LBGs \\
\hline
\hline
 & 
  $z=0$  & 
  $z=1.2$  & 
  $z=3.0$ \\
\hline
$n$  (h$_{100}^{3}$\,Mpc$^{-3}$) & 
  $6.0$\,$\cdot{}$\,10$^{-4}$ &
  \phantom{\,$\cdot{}$\,10$^{-3}$}($1.0\pm0.1$)\,$\cdot{}$\,10$^{-3}$ &
  \phantom{\,$\cdot{}$\,10$^{-3}$}($4.7\pm1.0$)\,$\cdot{}$\,10$^{-3}$ \\
$r_{0,{\rm g}}\,\, ({\rm h}_{100}^{-1}\,{\rm Mpc})$ & 
  $7.42\pm0.33$   &  
  $\phantom{0.5}9.5\pm0.5\phantom{9.5}$   &
  $3.8\pm0.3$  \\
$\gamma_{\rm g}$& 
  $-1.76\pm0.04\phantom{-}$   &  
  $-1.8\phantom{\pm-1.8}$   &
  $-1.60\pm0.15\phantom{-}$ \\
$\xi_{8,{g}}$   & 
  $0.88\pm0.07$ &  
  $1.36\pm0.13$ &  
  $0.30\pm0.08$ \\
b   & 
  $1.41\pm0.06$  &
  $2.7\pm0.1$  &
  $1.9\pm0.2$\\
\hline
\end{tabular}
\tablecomments{Number density ($n$), correlation length ($r_{0,g}$),
correlation function slope ($\gamma_g$), correlation amplitude at
8\,h$_{100}^{-1}$\,Mpc ($\xi_{8,{g}}$) and bias ($b$) for the observed
populations of galaxies discussed in the text. }
\end{center}
\end{table*}
\end{center}

\begin{center}
\begin{table*}
\begin{center}
\caption{\label{tab:masses}Characteristic Population Masses}
\begin{tabular}{ccccccc}%\hline
 \multicolumn{2}{c}{} &
 \multicolumn{1}{c}{$\alpha=0$} &
 \multicolumn{1}{c}{} &
 \multicolumn{3}{c}{$\alpha=0.8$} 
\\
\cline{3-3}\cline{5-7}%\\
Population & $z$ & $\Mmin$ (\hoh\,M$_{\odot}$)  &$\phantom{x}$ & $\Mmin$ (\hoh\,M$_{\odot}$) & $M_1$ (\hoh\,M$_{\odot}$) & $\Mbar$ (\hoh\,M$_{\odot}$)\\
\hline
\hline
gEs  & 0.0 & 1.7\,$\cdot{}$\,$10^{13}$ & & 2.7\,$\cdot{} 10^{12}$ & 4.4\,$\cdot{} 10^{13}$ & 8.9\,$\cdot{} 10^{13}$ \\
EROs & 1.2 & 1.6\,$\cdot{}$\,$10^{13}$ & & 9.1\,$\cdot{} 10^{12}$ & 3.9\,$\cdot{} 10^{12}$ & 3.8\,$\cdot{} 10^{13}$ \\
LBGs & 3.0 & 7.0\,$\cdot{}$\,$10^{10}$ & & 1.3\,$\cdot{} 10^{10}$ & 6.0\,$\cdot{} 10^{12}$ & 5.5\,$\cdot{} 10^{11}$ \\
\hline
\end{tabular}
\tablecomments{Characteristic mass values for each population
discussed in the text.  The $\Mmin$ for the $\alpha=0$ case (one
galaxy per DM halo) is given, as well as the $\Mmin$, $\Mn$ and
$\Mbar$ for the adopted $\alpha=0.8$ case.  These latter values
correspond to the points marked with double-circles in
Fig.~\ref{fig:invert}.}
\end{center}
\end{table*}
\end{center}

\end{document}